\documentclass[12pt]{report}
\usepackage[utf8]{inputenc}
\usepackage{blindtext}
\usepackage[letterpaper, total={8.5in, 11in}, margin=2in]{geometry}
\usepackage{ragged2e}
\usepackage{blindtext}
\usepackage{graphicx}
\usepackage{amsmath}
\usepackage{amssymb}
\usepackage{gensymb}
\usepackage{array}
\date{}

\begin{document}
\centering{\huge Improving SINR Performance Deploying IRS in 6G Wireless Networks\\
\vspace{24pt}
\large Mobasshir Mahbub, Raed M. Shubair}

\newpage

\RaggedRight{\textbf{\Large 1.\hspace{10pt} Introduction}}\\
\vspace{18pt}
\justifying{\noindent Over the past thirty years, wireless discoveries and methodologies, along with their prospective applications, have seen exponential progress and improvement. They include transmitter design as well as broadcast properties [1-18], THz interfacing and signal enhancement properties [19-31], as well as interior location strategies and problems [32-50].

Deploying miniature cells, aka small cells [51], within a very distracted network represents one of the simplest ways for meeting the demands of huge data traffic currently. To improve cellular services, such ground stations are put on-premises near resource-hungry equipment, generally in a smaller indoor/outdoor region. Furthermore, such base stations have higher coverage, ensuring adequate transmission. Because micro cells are located near to customers, a low-power transmission capacity (relative to large base stations) may ensure a better throughput to the gadgets [52].

The sixth version of wireless networking technologies aspires to build a data-driven, economically viable future enabled by near-instant, secure, unlimited, and environmentally friendly communication [53]. To accomplish this, industry and academics have established stringent performance criteria for security and reliability, speed, sensory capabilities, dependability, flexibility, and consuming energy. As a consequence, the 6G or sixth-generation cellular framework must be developed to meet growing requirements and standards, such as larger coverage, improved capacity, ubiquitous connection, and so on [54].

A planar metasurface made of an extensive number of reflecting components that has recently piqued academic attention due to its ability to significantly boost network energy as well as spectrum productivity by modifying wireless transmission settings termed as IRS. IRS segments with the appropriate phase shift may be able to relay the arriving wave [55]. By dynamically adjusting reflected signal communications, IRS creates constructive signal aggregation as well as harmful interference reduction at reception. As a result, the receiver's quality of experience (QoE) may improve.

As a consequence, the study evaluated the performance of normal miniature cellular connection with IRS-enhanced miniature cellular connectivity with regard to SINR in the case of services and applications.
}
\vspace{18pt}

\RaggedRight{\textbf{\Large 2.\hspace{10pt} Related Literature}}\\
\vspace{18pt}
\justifying{\noindent Wu et al. [56] investigated power regulation in the upstream for IRS-assisted connected devices under the restrictions of user-end level of performance (QoS). The study aimed to reduce users' overall power consumption by adjusting the phase transitions of IRS reflecting components and base station reception beamforming while keeping the SINR restriction in mind. Hao et al. [57] investigated an IRS-empowered multiuser MIMO system using orthogonal frequency-division multi-access (OFDMA) across the THz range. The scattered radio frequency, or RF, antenna design was used in the study to lower the network's energy use. The goal of the study is to optimize the sum-rate by enhancing hybrid beam formation simultaneously.}

\vspace{18pt}
\RaggedRight{\textbf{\Large 3.\hspace{10pt} Measurement Model}}\\
\vspace{12pt}

\justifying{\noindent 
Let’s consider, $P_t^D$ is the amount of power employed for transmission.
}

\vspace{12pt}
\RaggedRight{\textit{\large A.\hspace{10pt} Conventional Model}}\\
\vspace{12pt}
\justifying{\noindent For this instance, the power obtained by a typical miniature cell is depicted as follows (Eq. 1) [58], [59],
}
\vspace{6pt}
\begin{equation}
P_r^{DL}= \frac{\lambda L P_t^{DL}}{r^\alpha 16\pi^2 } 
\end{equation}
where $r = \sqrt{(x^{s}-x^{ue})^2+(y^{s}-y^{ue})^2+(z^{s}-z^{ue})^2}$. $\alpha$ is the fading level. $L$ represents a Rayleigh decay factor having an exponential form that follows a unit mean dispersion. $\lambda= c⁄f$  denotes the wavelength related to the carrier. $c$ is the velocity of the sunlight in $ms^{-1}$. The carrier resonance is represented by $f$ in Hz.

The equation beneath is used to compute the downstream SINR (Eq. 2),
\vspace{6pt}
\begin{equation}
S^{DL}= \frac{P_r^{DL}}{Intf.+n_w} 
\end{equation}
where $Intf.$ denotes the interfering signal. $n_w$ is noise.

\vspace{12pt}
\RaggedRight{\textit{\large B.\hspace{10pt} IRS-Empowered Model}}\\
\vspace{12pt}

\justifying{\noindent In the case of an IRS-assisted small cell for IoT components, the obtained power in the downstream is computed as follows (Eq. 2) [60],}

\begin{equation}
P_r^{DL(IRS)} = \frac{l_x w_y m^2 n^2 \lambda^2 G_T G_R G cos(\theta_T) cos(\theta_R) A^2}{64\pi^3(r_1 r_2)^2} P_t^{DL(IRS)}
\end{equation}

where $A$ is the numerical value that represents the refraction index. $\theta_T$ is the direction for transmitting and $\theta_R$ is the direction of reception. $l_x$ and $w_y$ are the length and width of IRS components. $G=\frac{4\pi l_x w_y}{\lambda^2}$  represents the IRS's dispersal gain. $m$ and $n$ represent the total quantity of pairs of transmitting and receiving elements. $G_T$ refers to the transmitting gain and $G_R$ refers to the recipient's gain. $r_1=  \sqrt{(x^{s}-x_i^{i} )^2+(y^{s}-y_i^{i} )^2+(z^{s}-z_i^{i} )^2}$ indicates the distance that lies between the signal's transmitter and the receiver at $(x^{s},y^{s},z^{s})$ coordinates and corresponding IRS at $(x_i^{i},y_i^{i},z_i^{i})$ coordinates. 

$r_2=  \sqrt{(x_i^{i}-x_n^{ue} )^2+(y_i^{i}-y_n^{ue} )^2+(z_i^{i}-z_n^{ue} )^2}$ is the separation between the IRS and the devices at $(x^{ue},y^{ue},z^{ue})$.

The subsequent equation yields the downstream SINR (Eq. 4),
\vspace{6pt}
\begin{equation}
S^{DL(IRS)}= \frac{P_r^{DL(IRS)}}{Intf.+n_w} 
\end{equation}

\vspace{18pt}

\RaggedRight{\textbf{\Large 4.\hspace{10pt} Numerical Results and Discussions}}\\
\vspace{12pt}
\justifying{\noindent The measurement findings produced through MATLAB-based modelling implementing the measurement concept are included in this portion of the article, as are remarks on the generated results.}

\vspace{12pt}
\RaggedRight{\textit{\large A.\hspace{10pt} Results for Conventional Model}}\\
\vspace{12pt}
\justifying{
Fig. 1 visualizes the measurement of downlink SINR for conventional micro cells.}

\vspace{12pt}
\centering{
\includegraphics[height=9.0cm, width=11.5cm]{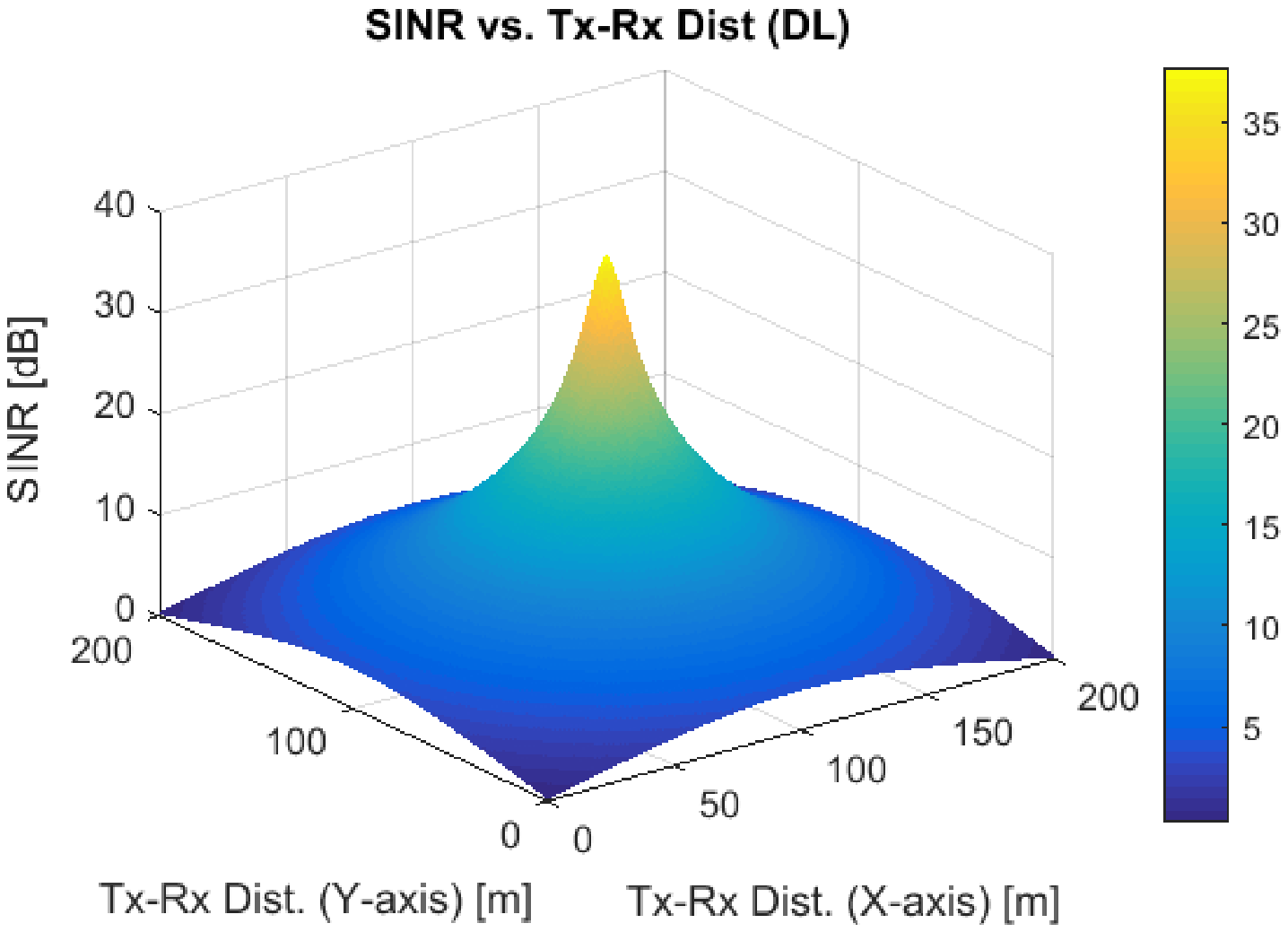}\par

Fig. 1. Transmitter-receiver separation distance vs. SINR (non-IRS DL)}

\justifying
According to the observation in Fig. 1, the devices located within 10 – 15 m will achieve around 24 dB SINR and the devices 40 m away from the base station will obtain around 14 dB SINR. The cell edge devices will obtain 0.4 - 4 dB of SINR approximately.

\vspace{12pt}
\RaggedRight{\textit{\large B.\hspace{10pt} Results for IRS-Assisted Model}}\\
\vspace{12pt}
\justifying{
Fig. 2 (a), (b), and (c) show the downlink SINR for the whole coverage region in terms of varied transmit-receive angles. Fig. 2 (d) visualizes the SINR placing the IRS 50 m away from the cell edge.}

\vspace{12pt}
\centering{
\includegraphics[height=9.0cm, width=11.5cm]{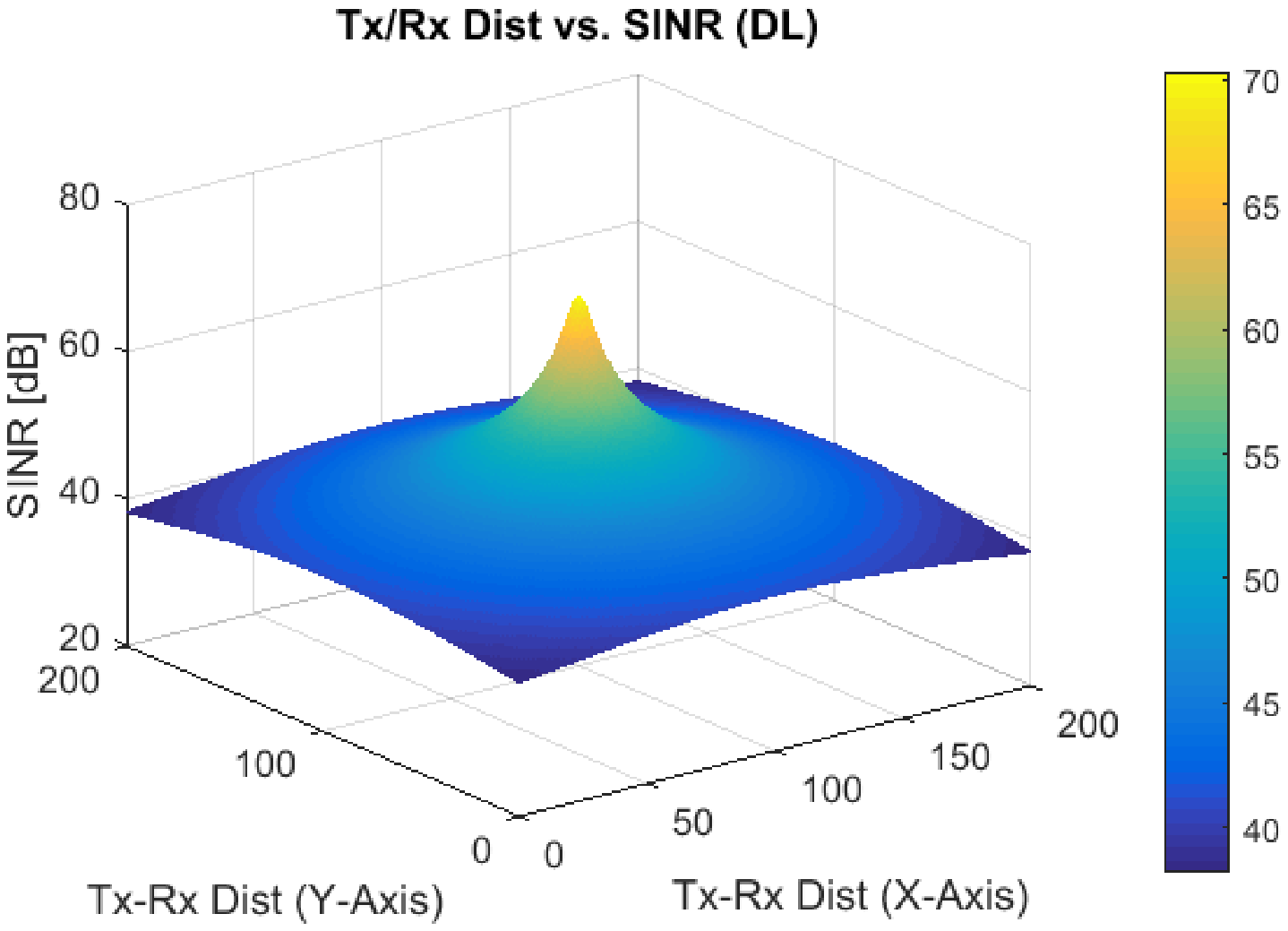}

(a)
}
 
\vspace{12pt}
\centering{
\includegraphics[height=9.0cm, width=11.5cm]{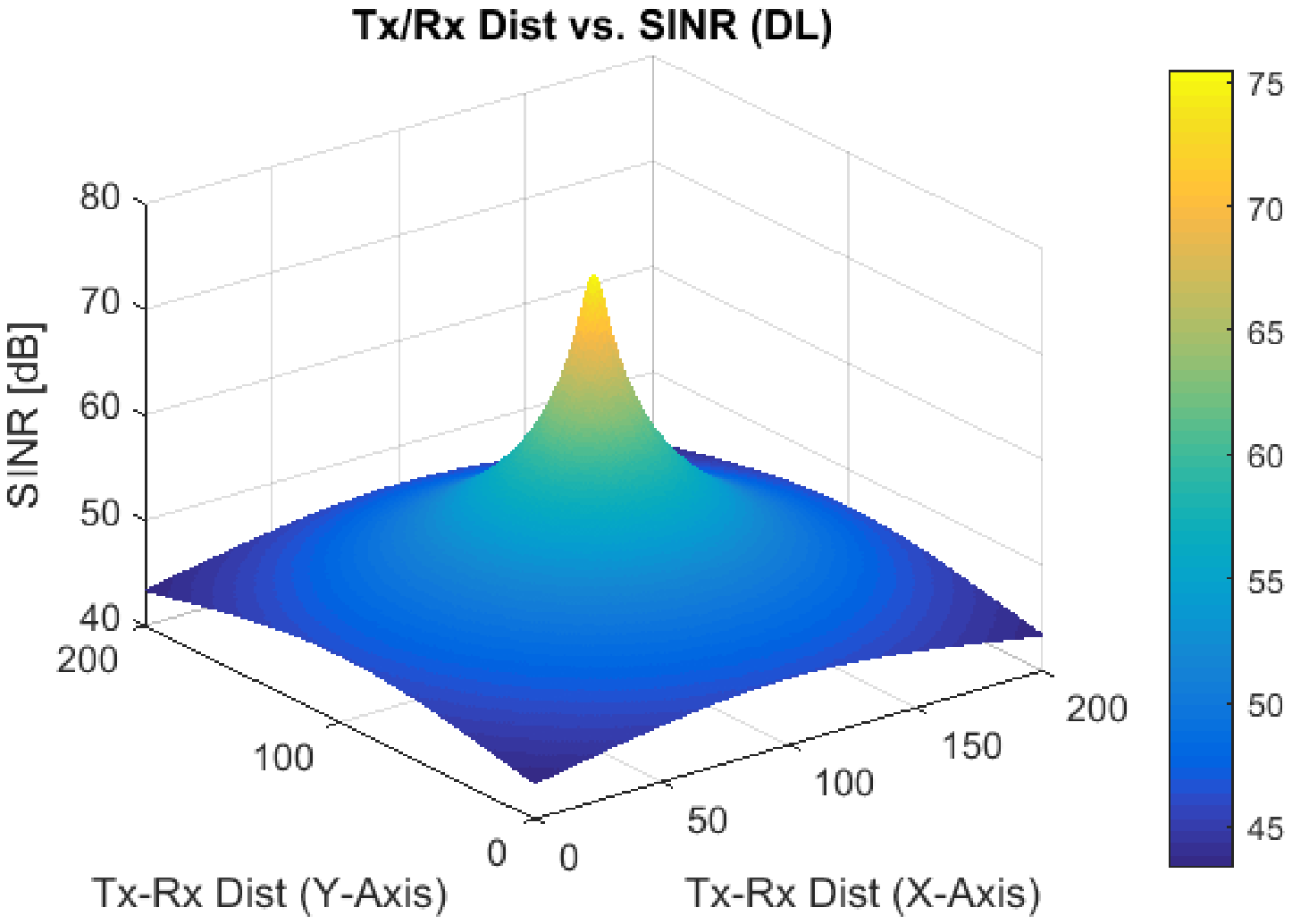}

(b)
}

\vspace{12pt}
\centering{
\includegraphics[height=9.0cm, width=11.5cm]{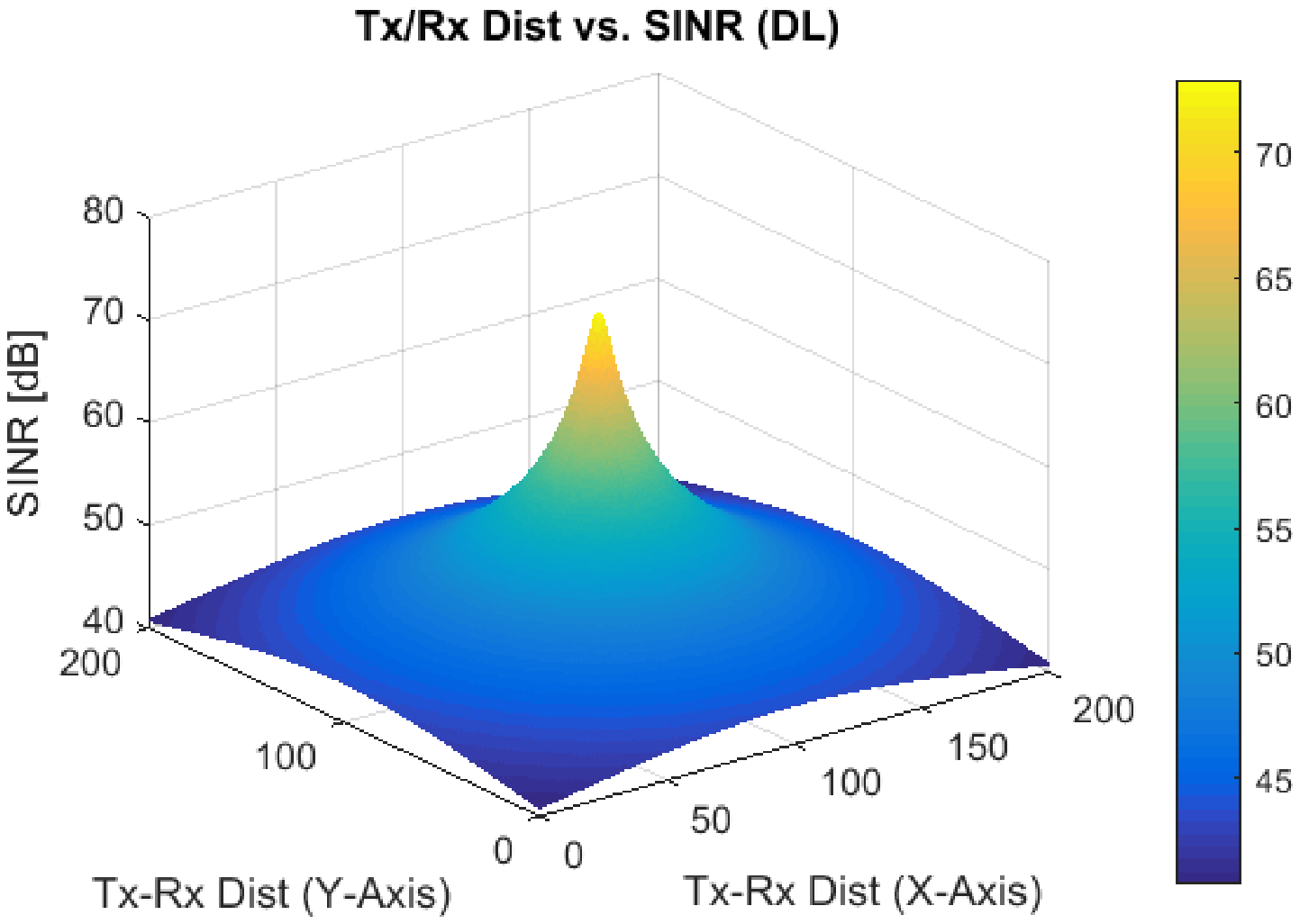}

(c)
}

\vspace{12pt}
\centering{
\includegraphics[height=9.0cm, width=11.5cm]{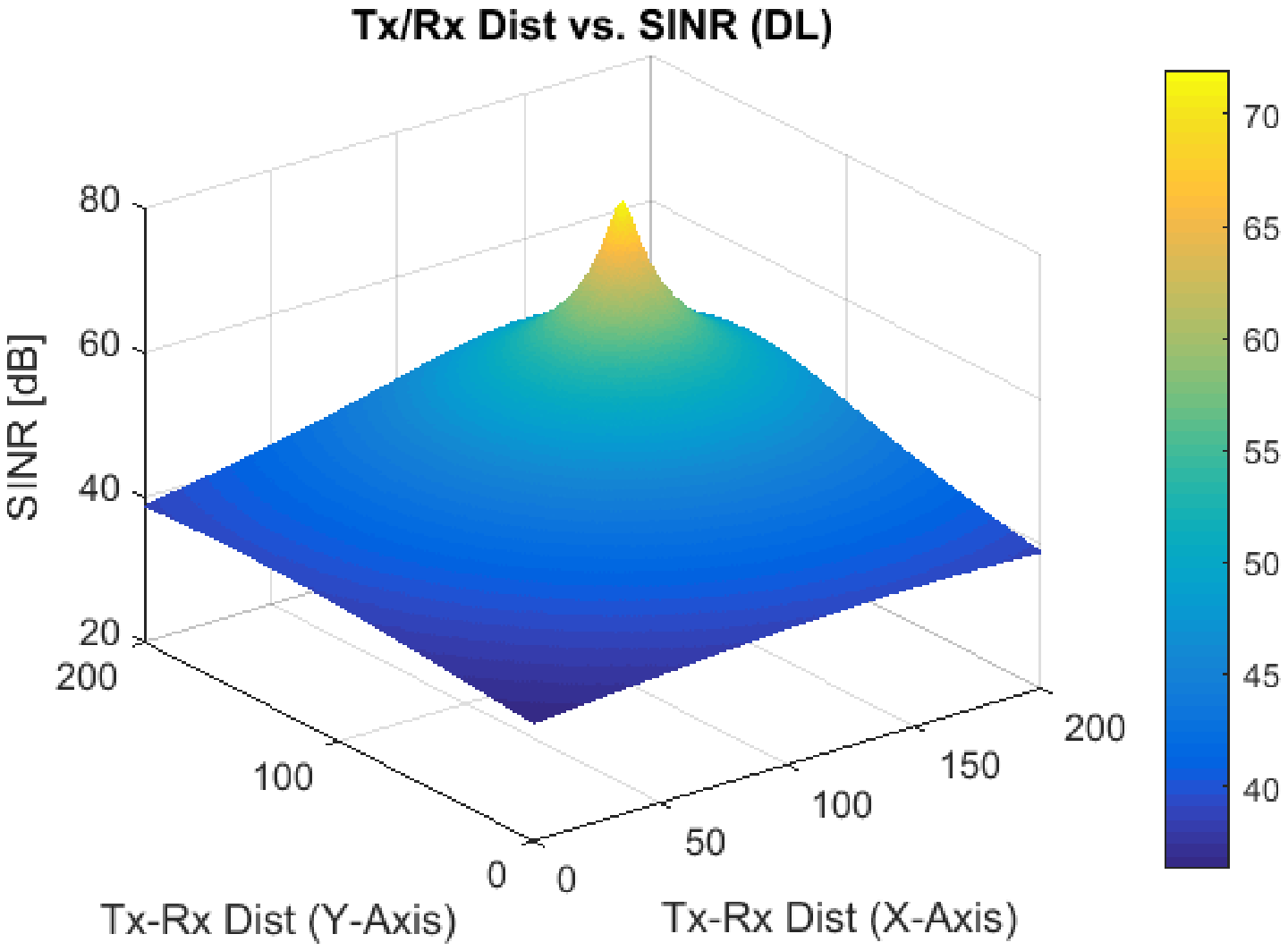}

(d)
}

\justifying{Fig. 2. (a) Transmitter-receiver separation vs. SINR ($\theta_t= 45\degree$, $\theta_r= 45\degree$), (b) Transmitter-receiver separation vs. SINR ($\theta_t= 60\degree$, $\theta_r= 60\degree$), (c) Transmitter-receiver separation vs. SINR ($\theta_t= 45\degree$, $\theta_r= 60\degree$), (d) Transmitter-receiver separation vs. SINR ($\theta_t= 60\degree$, $\theta_r= 60\degree$ and IRS placed at 50 m away from cell edge)

The observation of Fig. 2 (a)-(c) states that the $60\degree$ transmit-receive angles (from the base station to IRS and IRS to device) outperform other considered angles in terms of downlink received power. Afterward, $45\degree$ of transmitting and $60\degree$ of receiving angles perform better than other considered transmit-receive angles. In the case of the $60\degree$ transmit-receive angles, the devices located 15 m away from the IRS achieve around -25 dBm of received power, at 50 m the received power in downlink becomes -35 dBm, and at the cell edge, it is -41 to -44 dBm.

Inspecting Fig. 2 (a), (b), and (c) it is feasible that, $60\degree$ of transmit-receive angles outperforms other angles in terms of downlink SINR (maximum 75 dB and minimum 44 dB). When the devices are randomly distributed over a coverage region it is favorable to place IRS in the mid of the coverage region to offer a favorable coverage to all devices according to the comparison between Fig. 2 (b) and (d). The placement of IRS near the receiver may be favorable for a clustered group of devices located in a specific region.}

\vspace{18pt}

\RaggedRight{\textbf{\Large 5.\hspace{10pt} Conclusion}}\\
\vspace{12pt}

\justifying{\noindent The study aimed to improve coverage in the environment of upcoming 6G networks by installing IRS in a miniature cell. A literature assessment of related current research was conducted to give insight and identify research constraints or gaps. It developed a measurement approach for both traditional and IRS-assisted miniature cellular networks, which includes numerous equations for measuring SINR.}

\vspace{18pt}

\RaggedRight{\textbf{\Large References}}\\
\justifying{
1.	Mohamed I. AlHajri, Nazar T. Ali, and Raed M. Shubair. "Indoor localization for IoT using adaptive feature selection: A cascaded machine learning approach." IEEE Antennas and Wireless Propagation Letters 18, no. 11 (2019): 2306-2310.

2.	M. A. Al-Nuaimi, R. M. Shubair, and K. O. Al-Midfa. "Direction of arrival estimation in wireless mobile communications using minimum variance distortionless response." In The Second International Conference on Innovations in Information Technology (IIT’05), pp. 1-5. 2005.

3.	Ebrahim M. Al-Ardi, Raed M. Shubair, and Mohammed E. Al-Mualla. "Direction of arrival estimation in a multipath environment: An overview and a new contribution." Applied Computational Electromagnetics Society Journal 21, no. 3 (2006): 226.

4.	R. M. Shubair. "Robust adaptive beamforming using LMS algorithm with SMI initialization." In 2005 IEEE Antennas and Propagation Society International Symposium, vol. 4, pp. 2-5. IEEE, 2005.

5.	R. M. Shubair and Y. L. Chow. "A closed-form solution of vertical dipole antennas above a dielectric half-space." IEEE transactions on antennas and propagation 41, no. 12 (1993): 1737-1741.

6.	Ebrahim M. Al-Ardi, Raed M. Shubair, and Mohammed E. Al-Mualla. "Computationally efficient DOA estimation in a multipath environment using covariance differencing and iterative spatial smoothing." In 2005 IEEE International Symposium on Circuits and Systems, pp. 3805-3808. IEEE, 2005.

7.	E. M. Al-Ardi, Raed M. Shubair, and M. E. Al-Mualla. "Performance evaluation of direction finding algorithms for adapative antenna arrays." In 10th IEEE International Conference on Electronics, Circuits and Systems, 2003. ICECS 2003. Proceedings of the 2003, vol. 2, pp. 735-738. IEEE, 2003.

8.	M. I. AlHajri, N. Alsindi, N. T. Ali, and R. M. Shubair. "Classification of indoor environments based on spatial correlation of RF channel fingerprints." In 2016 IEEE international symposium on antennas and propagation (APSURSI), pp. 1447-1448. IEEE, 2016.

9.	Raed M. Shubair, and Ali Hakam. "Adaptive beamforming using variable step-size LMS algorithm with novel ULA array configuration." In 2013 15th IEEE International Conference on Communication Technology, pp. 650-654. IEEE, 2013.

10.	R. M. Shubair, and A. Merri. "Convergence of adaptive beamforming algorithms for wireless communications." In Proc. IEEE and IFIP International Conference on Wireless and Optical Communications Networks, pp. 6-8. 2005.

11.	Mohamed I. AlHajri, Nazar T. Ali, and Raed M. Shubair. "A machine learning approach for the classification of indoor environments using RF signatures." In 2018 IEEE Global Conference on Signal and Information Processing (GlobalSIP), pp. 1060-1062. IEEE, 2018.

12.	E. M. Al-Ardi, R. M. Shubair, and M. E. Al-Mualla. "Performance evaluation of the LMS adaptive beamforming algorithm used in smart antenna systems." In 2003 46th Midwest Symposium on Circuits and Systems, vol. 1, pp. 432-435. IEEE, 2003.

13.	Raed M. Shubair, Abdulrahman S. Goian, Mohamed I. AlHajri, and Ahmed R. Kulaib. "A new technique for UCA-based DOA estimation of coherent signals." In 2016 16th Mediterranean Microwave Symposium (MMS), pp. 1-3. IEEE, 2016.

14.	M. I. AlHajri, R. M. Shubair, L. Weruaga, A. R. Kulaib, A. Goian, M. Darweesh, and R. AlMemari. "Hybrid method for enhanced detection of coherent signals using circular antenna arrays." In 2015 IEEE International Symposium on Antennas and Propagation \& USNC/URSI National Radio Science Meeting, pp. 1810-1811. IEEE, 2015.

15.	R. M. Shubair, and A. Merri. "A convergence study of adaptive beamforming algorithms used in smart antenna systems." In 11th International Symposium on Antenna Technology and Applied Electromagnetics [ANTEM 2005], pp. 1-5. IEEE, 2005.

16.	E. M. Ardi, , R. M. Shubair, and M. E. Mualla. "Adaptive beamforming arrays for smart antenna systems: A comprehensive performance study." In IEEE Antennas and Propagation Society Symposium, 2004., vol. 3, pp. 2651-2654. IEEE, 2004.

17.	Raed M. Shubair, and Hadeel Elayan. "Enhanced WSN localization of moving nodes using a robust hybrid TDOA-PF approach." In 2015 11th International Conference on Innovations in Information Technology (IIT), pp. 122-127. IEEE, 2015.

18.	M. I. AlHajri, N. T. Ali, and R. M. Shubair. "2.4 ghz indoor channel measurements data set." UCI Machine Learning Repository (2018).

19.	Mohamed I. AlHajri, Raed M. Shubair, and Marwa Chafii. "Indoor Localization Under Limited Measurements: A Cross-Environment Joint Semi-Supervised and Transfer Learning Approach." In 2021 IEEE 22nd International Workshop on Signal Processing Advances in Wireless Communications (SPAWC), pp. 266-270. IEEE, 2021.

20.	Raed M. Shubair and Hadeel Elayan. "In vivo wireless body communications: State-of-the-art and future directions." In 2015 Loughborough Antennas \& Propagation Conference (LAPC), pp. 1-5. IEEE, 2015.

21.	Hadeel Elayan, Raed M. Shubair, and Asimina Kiourti. "Wireless sensors for medical applications: Current status and future challenges." In 2017 11th European Conference on Antennas and Propagation (EUCAP), pp. 2478-2482. IEEE, 2017.

22.	Hadeel Elayan, Raed M. Shubair, Josep Miquel Jornet, and Raj Mittra. "Multi-layer intrabody terahertz wave propagation model for nanobiosensing applications." Nano communication networks 14 (2017): 9-15.

23.	Samar Elmeadawy and Raed M. Shubair. "6G wireless communications: Future technologies and research challenges." In 2019 international conference on electrical and computing technologies and applications (ICECTA), pp. 1-5. IEEE, 2019.

24.	Maryam AlNabooda, Raed M. Shubair, Nadeen R. Rishani, and GhadahAldabbagh. "Terahertz spectroscopy and imaging for the detection and identification of illicit drugs." 2017 Sensors networks smart and emerging technologies (SENSET) (2017): 1-4.

25.	Hadeel Elayan, Raed M. Shubair, Akram Alomainy, and Ke Yang. "In-vivo terahertz em channel characterization for nano-communications in wbans." In 2016 IEEE International Symposium on Antennas and Propagation (APSURSI), pp. 979-980. IEEE, 2016.

26.	Hadeel Elayan, Raed M. Shubair, and Josep M. Jornet. "Bio-electromagnetic thz propagation modeling for in-vivo wireless nanosensor networks." In 2017 11th European Conference on Antennas and Propagation (EuCAP), pp. 426-430. IEEE, 2017.

27.	Hadeel Elayan, Raed M. Shubair, and Josep M. Jornet. "Characterising THz propagation and intrabody thermal absorption in iWNSNs." IET Microwaves, Antennas \& Propagation 12, no. 4 (2018): 525-532.

28.	Dana Bazazeh, Raed M. Shubair, and Wasim Q. Malik. "Biomarker discovery and validation for Parkinson's Disease: A machine learning approach." In 2016 International Conference on Bio-engineering for Smart Technologies (BioSMART), pp. 1-6. IEEE, 2016.

29.	Hadeel Elayan, Cesare Stefanini, Raed M. Shubair, and Josep M. Jornet. "Stochastic noise model for intra-body terahertz nanoscale communication." In Proceedings of the 5th ACM International Conference on Nanoscale Computing and Communication, pp. 1-6. 2018.

30.	Hadeel Elayan, Hadeel, and Raed M. Shubair. "Towards an Intelligent Deployment of Wireless Sensor Networks." In Information Innovation Technology in Smart Cities, pp. 235-250. Springer, Singapore, 2018.
31.	Hadeel Elayan, Raed M. Shubair, Josep M. Jornet, Asimina Kiourti, and Raj Mittra. "Graphene-Based Spiral Nanoantenna for Intrabody Communication at Terahertz." In 2018 IEEE International Symposium on Antennas and Propagation \& USNC/URSI National Radio Science Meeting, pp. 799-800. IEEE, 2018.

32.	Abdul Karim Gizzini, Marwa Chafii, Shahab Ehsanfar, and Raed M. Shubair. "Temporal Averaging LSTM-based Channel Estimation Scheme for IEEE 802.11 p Standard." arXiv preprint arXiv:2106.04829 (2021).

33.	Ahmed A. Ibrahim,  JanMachac, and Raed M. Shubair. "Compact UWB MIMO antenna with pattern diversity and band rejection characteristics." Microwave and Optical Technology Letters 59, no. 6 (2017): 1460-1464.

34.	M. Saeed Khan, A-D. Capobianco, Sajid M. Asif, Adnan Iftikhar, Benjamin D. Braaten, and Raed M. Shubair. "A pattern reconfigurable printed patch antenna." In 2016 IEEE International Symposium on Antennas and Propagation (APSURSI), pp. 2149-2150. IEEE, 2016.

35.	Muhammad Saeed Khan, Adnan Iftikhar, Antonio‐Daniele Capobianco, Raed M. Shubair, and Bilal Ijaz. "Pattern and frequency reconfiguration of patch antenna using PIN diodes." Microwave and Optical Technology Letters 59, no. 9 (2017): 2180-2185.

36.	Muhammad Saeed Khan, Adnan Iftikhar, Raed M. Shubair, Antonio-D. Capobianco, Benjamin D. Braaten, and Dimitris E. Anagnostou. "Eight-element compact UWB-MIMO/diversity antenna with WLAN band rejection for 3G/4G/5G communications." IEEE Open Journal of Antennas and Propagation 1 (2020): 196-206.

37.	Raed M. Shubair, Amna M. AlShamsi, Kinda Khalaf, and Asimina Kiourti. "Novel miniature wearable microstrip antennas for ISM-band biomedical telemetry." In 2015 Loughborough Antennas \& Propagation Conference (LAPC), pp. 1-4. IEEE, 2015.

38.	Ala Eldin Omer, George Shaker, Safieddin Safavi-Naeini, Georges Alquié, Frédérique Deshours, Hamid Kokabi, and Raed M. Shubair. "Non-invasive real-time monitoring of glucose level using novel microwave biosensor based on triple-pole CSRR." IEEE Transactions on Biomedical Circuits and Systems 14, no. 6 (2020): 1407-1420.

39.	M. S. Khan, F. Rigobello, Bilal Ijaz, E. Autizi, A. D. Capobianco, R. Shubair, and S. A. Khan. "Compact 3‐D eight elements UWB‐MIMO array." Microwave and Optical Technology Letters 60, no. 8 (2018): 1967-1971.

40.	R. Karli, H. Ammor, R. M. Shubair, M. I. AlHajri, and A. Hakam. "Miniature Planar Ultra-Wide-Band Microstrip Patch Antenna for Breast Cancer Detection." Skin 1 (2016): 39.

41.	Muhammad Saeed Khan, Adnan Iftikhar, Raed M. Shubair, Antonio-Daniele Capobianco, Sajid Mehmood Asif, Benjamin D. Braaten, and Dimitris E. Anagnostou. "Ultra-compact reconfigurable band reject UWB MIMO antenna with four radiators." Electronics 9, no. 4 (2020): 584.

42.	Amjad Omar, Maram Rashad, Maryam Al-Mulla, Hussain Attia, Shaimaa Naser, Nihad Dib, and Raed M. Shubair. "Compact design of UWB CPW-fed-patch antenna using the superformula." In 2016 5th International Conference on Electronic Devices, Systems and Applications (ICEDSA), pp. 1-4. IEEE, 2016.

43.	Hari Shankar Singh, SachinKalraiya, Manoj Kumar Meshram, and Raed M. Shubair. "Metamaterial inspired CPW‐fed compact antenna for ultrawide band applications." International Journal of RF and Microwave Computer‐Aided Engineering 29, no. 8 (2019): e21768.

44.	Omar Masood Khan, Qamar Ul Islam, Raed M. Shubair, and Asimina Kiourti. "Novel multiband Flamenco fractal antenna for wearable WBAN off-body communication applications." In 2018 International Applied Computational Electromagnetics Society Symposium (ACES), pp. 1-2. IEEE, 2018.

45.	Yazan Al-Alem, Ahmed A. Kishk, and Raed M. Shubair. "Enhanced wireless interchip communication performance using symmetrical layers and soft/hard surface concepts." IEEE Transactions on Microwave Theory and Techniques 68, no. 1 (2019): 39-50.

46.	Yazan Al-Alem, Ahmed A. Kishk, and Raed M. Shubair. "One-to-two wireless interchip communication link." IEEE Antennas and Wireless Propagation Letters 18, no. 11 (2019): 2375-2378.

47.	Asimina Kiourti, and Raed M. Shubair. "Implantable and ingestible sensors for wireless physiological monitoring: a review." In 2017 IEEE International Symposium on Antennas and Propagation \& USNC/URSI National Radio Science Meeting, pp. 1677-1678. IEEE, 2017.

48.	Yazan Al-Alem, Raed M. Shubair, and Ahmed Kishk. "Clock jitter correction circuit for high speed clock signals using delay units a nd time selection window." In 2016 16th Mediterranean Microwave Symposium (MMS), pp. 1-3. IEEE, 2016.

49.	Yazan Al-Alem, Ahmed A. Kishk, and Raed Shubair. "Wireless chip to chip communication link budget enhancement using hard/soft surfaces." In 2018 IEEE Global Conference on Signal and Information Processing (GlobalSIP), pp. 1013-1014. IEEE, 2018.

50.	Yazan Al-Alem, Yazan, Ahmed A. Kishk, and Raed M. Shubair. "Employing EBG in Wireless Inter-chip Communication Links: Design and Performance." In 2020 IEEE International Symposium on Antennas and Propagation and North American Radio Science Meeting, pp. 1303-1304. IEEE, 2020.

51.	S. Rajoria, A. Trivedi, and W. W. Godfrey, “A comprehensive survey: Small cell meets massive MIMO,” Physical Communication, vol. 26, pp. 40-49, February 2018.

52.	M. H. Alsharif, R. Nordin, M. M. Shakir, “Small Cells Integration with the Macro-Cell Under LTE Cellular Networks and Potential Extension for 5G,” Journal of Electrical Engineering and Technology, vol. 14, pp. 2455–2465, April 2019.

53.	L. Qiao et al., “A survey on 5G/6G, AI, and Robotics,” Computers \& Electrical Engineering, vol. 95, Oct. 2021.

54.	X. You, et al., “Towards 6G wireless communication networks: vision, enabling technologies, and new paradigm shifts,” Science China Information Sciences, vol. 64, Nov. 2020.

55.	F. Guo, F. R. Yu, H. Zhang, X. Li, H. Ji and V. C. M. Leung, "Enabling Massive IoT Toward 6G: A Comprehensive Survey," in IEEE Internet of Things Journal, vol. 8, no. 15, pp. 11891-11915, Aug. 2021.

56.	J. Wu and B. Shim, "Power Minimization of Intelligent Reflecting Surface-Aided Uplink IoT Networks," 2021 IEEE Wireless Communications and Networking Conference (WCNC), 2021, pp. 1-6.

57.	W. Hao et al., "Robust Design for Intelligent Reflecting Surface-Assisted MIMO-OFDMA Terahertz IoT Networks," in IEEE Internet of Things Journal, vol. 8, no. 16, pp. 13052-13064, 15 Aug.15, 2021.

58.	T. Mir, L. Dai, Y. Yang, W. Shen and B. Wang, "Optimal FemtoCell Density for Maximizing Throughput in 5G Heterogeneous Networks under Outage Constraints," 2017 IEEE 86th Vehicular Technology Conference (VTC-Fall), Toronto, ON, Canada, 2017, pp. 1-5.

59.	M. Mozaffari, W. Saad, M. Bennis and M. Debbah, "Optimal Transport Theory for Cell Association in UAV-Enabled Cellular Networks," in IEEE Communications Letters, vol. 21, no. 9, pp. 2053-2056, Sept. 2017.

60.	W. Tang et al., "Wireless Communications With Reconfigurable Intelligent Surface: Path Loss Modeling and Experimental Measurement," in IEEE Transactions on Wireless Communications, vol. 20, no. 1, pp. 421-439, Jan. 2021.
}

\end{document}